\documentclass[aps,prd, notitlepage, onecolumn,superscriptaddress,floatfix,letterpaper,nofootinbib, longbibliography]{revtex4-1}

\usepackage{amsmath, amsthm, amssymb,slashed,mathtools,tabu}
\usepackage{pifont}

\usepackage[usenames, dvipsnames]{color}
\usepackage[svgnames]{xcolor}
\usepackage[colorlinks,citecolor=RoyalBlue, urlcolor=RoyalBlue, linkcolor=RoyalBlue ]{hyperref} 

\usepackage[normalem]{ulem}

\newcommand{\cblue}[1]{\textcolor{black}{#1}}

\usepackage{booktabs}

\newcommand{\ccblue}[1]{\textcolor{black}{#1}}

\definecolor{mygray}{gray}{0.6}

\usepackage{upgreek}
\usepackage{bbm}

\newenvironment{myfont}[2][]{\csname#2\endcsname[#1]}{}

\usepackage{slashed}
\usepackage[makeroom]{cancel}
\usepackage[normalem]{ulem}
\usepackage{soul}
\newcommand{\stkout}[1]{\ifmmode\text{\sout{\ensuremath{#1}}}\else\sout{#1}\fi}

\usepackage{sseq}
\usepackage[all,cmtip]{xy}
\usepackage{tikz-cd}
\usepackage{tikz}
\usetikzlibrary{matrix}
\usetikzlibrary{decorations.markings}
\usetikzlibrary{tikzmark,decorations.pathreplacing,positioning}
\usepackage{amsfonts}

\newcommand{\bea}{\begin{eqnarray}}
\newcommand{\eea}{\end{eqnarray}}
\def\be{\begin{equation}}
\def\ee{\end{equation}}

\newcommand{\e}{\hspace{1pt}\mathrm{e}}

\newcommand{\ii}{\hspace{1pt}\mathrm{i}\hspace{1pt}}

\definecolor{red}{rgb}{1,0,0}
\definecolor{blue}{rgb}{0,0,1}
\definecolor{dblue}{rgb}{0,0,0.4}
\definecolor{green}{rgb}{0,1,0}
\definecolor{black}{rgb}{0,0,0}
\definecolor{white}{rgb}{1,1,1}

\definecolor{brn}{rgb}{.8,.4,.0}
\definecolor{redo}{rgb}{1,.5,.0}
\definecolor{ddgrn}{rgb}{0,0.4,0}
\definecolor{dgrn}{rgb}{0,0.55,0}
\definecolor{dbl}{rgb}{0,0,0.5}

\usepackage[bbgreekl]{mathbbol}
\usepackage{amscd}

\newcommand{\Z}{\mathbb{Z}}

\newcommand{\dd}{\hspace{1pt}\mathrm{d}}

\newcommand{\Refe}[1]{Ref.~[\onlinecite{#1}]}

\newcommand{\eq}[1]{(\ref{#1})}

\newcommand{\prt}{\partial}

\newcommand{\bpm}{\begin{pmatrix}}
\newcommand{\epm}{\end{pmatrix}}
\newcommand{\bmm}{\begin{matrix}}
\newcommand{\emm}{\end{matrix}}

\newcommand{\cD}{ {\cal D} }

\newcommand{\cL}{ {\cal L} }

\def\Z{{\mathbb{Z}}}

\def \H{\operatorname{H}}

\def \Z{\mathbb{Z}}

\newcommand {\emptycomment}[1]{}

\def\TP{\mathrm{TP}}

\def\B{\mathrm{B}}

\newcommand{\Spin}{{\rm Spin}}
\newcommand{\U}{{\rm U}}
\newcommand{\SU}{{\rm SU}}
\newcommand{\PSU}{{\rm PSU}}

\def\bZ{{\mathbf{Z}}}
\usepackage{centernot}

\usepackage{enumitem} 
\usepackage{mathtools,amssymb,varwidth}

\usepackage{datetime}

\newcommand{\rL}{{\rm L}}
\newcommand{\rR}{{\rm R}}

\def\ra{\mathrm{a}}

\newcommand{\SM}{{\rm SM}}
\newcommand{\rmod}{\;{\rm mod}\;}

\begin{document}


\title{Proton Stability: From the Standard Model to Beyond Grand Unification
}

\author{Juven Wang 
}
\email[]{jw@cmsa.fas.harvard.edu}
\affiliation{Center of Mathematical Sciences and Applications, Harvard University, MA 02138, USA}

\author{Zheyan Wan}
\affiliation{Yau Mathematical Sciences Center, Tsinghua University, Beijing 100084, China}

\author{Yi-Zhuang You}
\affiliation{Department of Physics, University of California, San Diego, CA 92093, USA}

\begin{abstract}

A proton is known for its longevity, but what is its lifetime? 
While many Grand Unified Theories predict the proton decay with a finite lifetime,
we show that the Standard Model (SM) and some versions of Ultra Unification (which replace sterile neutrinos 
with new exotic gapped/gapless sectors, e.g., topological or conformal field theory under global anomaly cancellation constraints) 
with a discrete baryon plus lepton symmetry
permit a stable proton.
For the 4d SM with Lie group
$G_{\SM_q} \equiv \frac{\SU(3) \times   \SU(2) \times \U(1)_{\tilde Y}}{\Z_q}$ of $q=1,2,3,6$ and $N_f$ families of 15 or 16 Weyl fermions, 
in addition to the continuous baryon minus lepton $\U(1)_{\bf B - L}$ symmetry,
there is also a compatible discrete baryon plus lepton $\Z_{2N_f, \bf B + L}$ symmetry.
The $\Z_{2N_f, \bf B + L}$ is discrete due to the ABJ anomaly under the BPST SU(2) instanton.
Although both $\U(1)_{\bf B - L}$ and $\Z_{2N_f, \bf B + L}$ symmetries are anomaly-free under the dynamically gauged $G_{\SM_q}$,
it is important to check whether they have mixed anomalies 
with the gravitational background field (spacetime diffeomorphism under Spin group rotation)
and higher symmetries (whose charged objects are Wilson electric or 't Hooft magnetic line operators) of SM.
We can also replace the $\U(1)_{\bf B - L}$ with a discrete variant $\Z_{4,X}$ for $X \equiv 5({\bf B - L})-\frac{2}{3} {\tilde Y}$ of electroweak hypercharge ${\tilde Y}$.
We explore a systematic classification of candidate perturbative local and nonperturbative global anomalies
of the 4d SM, including all these gauge and gravitational backgrounds, via a cobordism theory, which controls the SM's deformation class.  
We discuss the proton stability of the SM and Ultra Unification in the presence of
discrete ${\bf B + L}$ symmetry protection, 
in particular $(\U(1)_{\bf B - L} \times \Z_{2N_f,\bf B + L})/{\Z_2^{F}}$ or  $(\Z_{4,X} \times \Z_{2N_f, \bf B + L})/{\Z_2^{F}}$ with the fermion parity $\Z_2^{F}$. 
%

\end{abstract}


\maketitle

\section{Introduction and Summary}

Proton is known for its longevity, but what is its lifetime?  
The observed universe is about $10^{10}$ years old,
while the proton mean lifetime is experimentally tested to be more than $10^{30} \sim 10^{34}$ years \cite{PDG2020, NathFileviez0601023, Senjanovic0912.5375}.
To date, all experiments
that attempt to go beyond the Standard Model (SM) \cite{Glashow1961trPartialSymmetriesofWeakInteractions, Weinberg1967tqSMAModelofLeptons, Salam1968}
to observe proton decay predicted by Grand Unified Theories (GUTs) 
\cite{Pati1974yyPatiSalamLeptonNumberastheFourthColor, Georgi1974syUnityofAllElementaryParticleForces, GeorgiQuinnWeinberg1974}
have not yet succeeded.
This motivates us to ask the following questions
and seek their resolutions:

1. Are there alternative routes to test GUT, other than conventionally seeking
GUT as an effective field theory that appeared at a higher energy unification?

2. Can the proton be stable with an infinite lifetime?
What mechanism protects the proton from decay?

For the first question, \Refe{Wang2106.16248, WangYou2111.10369GEQC, WangWanYou2112.14765, YouWang2202.13498}
recently suggested that instead of only increasing the energy scale from the SM to higher energy looking for the GUT structure 
(imagining tuning the energy scale along a \emph{vertical} axis in a phase diagram),
we can indeed move from the Standard Model (SM) vacuum to neighbor GUT vacua
via quantum phase transitions \cite{subirsachdev2011book}
(imagining tuning the vacuum changing parameters along a \emph{horizontal} axis in a quantum phase diagram at zero temperature).
In \Refe{Wang2106.16248, WangYou2111.10369GEQC, WangWanYou2112.14765, YouWang2202.13498} viewpoint,
the 4d SM and other GUTs are in the same \emph{deformation class} of quantum field theories \cite{NSeiberg-Strings-2019-talk},
labeled by $(G, \bZ_{5})$, the symmetry $G$ and its anomaly $\bZ_{5}$:
In the limit when internal symmetry is weakly coupled or ungauged,
the SM and GUTs could be labeled by an enlarged spacetime-internal symmetry group $G$
and a certain 't Hooft anomaly \cite{tHooft1979ratanomaly} of the symmetry $G$.
In a modern quantum field theory (QFT) language, 't Hooft anomaly of the symmetry $G$ in $d$d spacetime
is specified via the anomaly inflow \cite{1984saCallanHarvey, Witten2019bou1909.08775}
by a $(d+1)$d $G$-symmetric invertible topological quantum field theory (TQFT) denoted 
as a cobordism invariant $\bZ_{d+1}$ \cite{2016arXiv160406527F}.
Thus, we can tune the SM via quantum phase transitions
to the neighbor GUT phases that necessarily allow proton decays.
Theoretically, those quantum vacua tuning parameters can be the sign-flipping of the $r$ coefficient 
of the GUT-Higgs potential
${\rm U}(\Phi_{{\rm GUT}}^{})=
( r_{}^{} (\Phi_{{\rm GUT}}^{})^2 +\lambda_{}^{} (\Phi_{{\rm GUT}}^{})^4)$ or the sign flipping of the fermion mass.

For the second question, the proton longevity may 
be protected by subtle symmetries of QFT vacuum (low-energy ground states).
Those subtle symmetries, if found,
may not be accidental, but be exact ---
they seem global symmetries at SM energy scales,
but they should be dynamical gauge symmetries  \cite{KraussWilczekPRLDiscrete1989}
when approaching the Planck or quantum gravity scales,
due to ``no global symmetries in quantum gravity'' reasoning (see a recent overview \cite{HarlowOoguri1810.05338}).

Those subtle symmetries likely are discrete symmetries, 
subject to the nontrivial check of their anomaly matching or cancellation that we will perform.
These discrete symmetries and their anomalies were investigated in the past 
(e.g., \cite{Ibanez1991hvRossPLB, Ibanez1992NPB, BanksDine1991xj9109045, Csaki1997awMurayama9710105}),
they are famous for having potential \emph{nonperturbative global anomalies} \cite{Witten1985xe}
(classified by finite abelian group $\mathbb{Z}_{\rm n}$, 
detectable via large gauge/diffeomorphism transformations that \emph{cannot} be continuously deformed from the identity),
in contrast to the familiar \emph{perturbative local anomalies} 
(classified by integer $\Z$ classes,
detectable via infinitesimal gauge/diffeomorphism transformations continuously deformable from the identity,
captured by Feynman diagram calculations).
These \emph{local} and \emph{global} anomalies include \emph{gauge, gravitational} or \emph{mixed gauge-gravitational} anomaly types depending 
on whether their path integral non-invariance is due to gauge or gravitational background fields.
But only recently, thanks to the development of cobordism group classification of anomalies, 
these global anomalies become systematically computable \cite{2016arXiv160406527F, GarciaEtxebarriaMontero2018ajm1808.00009,Hsieh2018ifc1808.02881, WanWang2018bns1812.11967, DavighiGripaiosLohitsiri2019rcd1910.11277, WW2019fxh1910.14668}.

Discrete symmetries and their global anomalies can drastically challenge the paradigm that we used to think of QFT vacuum.
For example, if the baryon minus lepton ${\mathbf{B} - \mathbf{L}}$ vector symmetry or 
more precisely $X \equiv 5({ \mathbf{B}-  \mathbf{L}})-\frac{2}{3} {\tilde Y}$ chiral symmetry \cite{Wilczek1979hcZee, WilczekZeePLB1979}
(with the integer quantized electroweak hypercharge ${\tilde Y}$) is a discrete $\Z_{4,X}$ symmetry,
although the $\Z_{4,X}$ has no local anomalies,  the $\Z_{4,X}$ imposes various global anomaly cancellation conditions 
\cite{GarciaEtxebarriaMontero2018ajm1808.00009, Hsieh2018ifc1808.02881, GuoJW1812.11959, WW2019fxh1910.14668}.
In particular, a $\Z_{16}$ class global anomaly of the mixed gauge-gravity type 
(variation on the $\Z_{4,X}$ gauge field and the Spin spacetime diffeomorphism)
implies that $15N_f$ Weyl fermion SM alone cannot cancel the $\Z_{16}$ global anomaly 
--- its anomaly cancellation requires introducing either the 16th Weyl fermion (the right-handed neutrino),
or 4d noninvertible TQFT or interacting conformal field theory (CFT), or 5d invertible TQFT, 
or breaking the $\Z_{4,X}$ down to fermion parity $\Z_2^{F}$, etc \cite{JW2006.16996, JW2008.06499, JW2012.15860}.
In other words, rephrasing in terms of quantum phases of QFT language, 
the QFT vacuum beyond the SM (BSM) could be more quantumly entangled than the Landau-Ginzburg old paradigm.
These possible exotic BSM phases are analogous to many exotic quantum phases explored in the contemporary condensed matter community \cite{UQM}.
The SM together with those exotic BSM phases constrained by nonperturbative global anomaly cancellation is called Ultra Unification (UU) \cite{JW2006.16996, JW2008.06499, JW2012.15860}. 

Another motivation 
for our present work is expanding the exploration of the deformation class of SM \cite{WangWanYou2112.14765}.
In \Refe{WangWanYou2112.14765}, we explored the deformation of SM to GUTs.
We had included the spacetime symmetry (Spin group), 
the internal symmetry 
(Lie algebra $su(3) \times  su(2) \times u(1)_{\tilde Y}$, 
and four compatible versions Lie groups
$G_{\SM_q} \equiv \frac{\SU(3) \times   \SU(2) \times \U(1)_{\tilde Y}}{\Z_q}$,  with $q=1,2,3,6$ \cite{Tong2017oea1705.01853}),
and a continuous or discrete ${\mathbf{B} - \mathbf{L}}$ like symmetry of the SM.
In \Refe{WangWanYou2112.14765}, 
we had left the inclusion of the discrete ${\mathbf{B} + \mathbf{L}}$ vector symmetry for future work.
(The $\U(1)_{\mathbf{B} + \mathbf{L}}$ is explicitly broken down to $\Z_{2N_f \#, \mathbf{B} + \mathbf{L}}$ 
due to the SU(2) instanton \cite{BelavinBPST1975, tHooft1976ripPRL, tHooft1976instanton}
by the Adler-Bell-Jackiw (ABJ) anomaly \cite{Adler1969gkABJ, Bell1969tsABJ},
where $N_f$ is the family number
with an extra $q$-dependent factor denoted as $\#$, see \cite{AnberPoppitz2110.02981}.)
Previously we had excluded the ${\mathbf{B} + \mathbf{L}}$ because it is not a symmetry for many GUTs. 
Now the discrete ${\mathbf{B} + \mathbf{L}}$, allowed in SM but disobeyed by GUTs, 
gives us the exact opportunity to distinguish the SM from other GUTs.
Our present work means to fill this gap left in \cite{WangWanYou2112.14765}
to include 
the discrete ${\mathbf{B} + \mathbf{L}}$ symmetry, 
examining the ${\mathbf{B} + \mathbf{L}}$ anomaly, and its SM deformation class.
We will see that the discrete ${\mathbf{B} + \mathbf{L}}$ is a good symmetry for SM and some versions of UU,
such that it implies the proton stability in those models.
Remarkably, recently Ref.~\cite{ByaktiGhoshSharmaJHEPZN1707.03837, 
KorenProtonStability2204.01741} also emphasize that the discrete ${\mathbf{B} + \mathbf{L}}$ 
can avoid the proton decay.

\section{Revisit the Standard Model 
}

Now we revisit the Standard Model (SM) and its symmetry, then explicitly derive the discrete ${\mathbf{B} + \mathbf{L}}$ symmetry.
SM is a 4d chiral gauge theory of local Lie algebra $su(3) \times  su(2) \times u(1)_{\tilde Y}$
coupling to $N_f=3$ families of 15 or 16 Weyl fermions (written as a left-handed 15 or 16 multiplet $\psi_L$)
in the following representation
\bea \label{eq:SMrep}
({\psi_L})_{\rm I} =
( \bar{d}_R \oplus {l}_L  \oplus q_L  \oplus \bar{u}_R \oplus   \bar{e}_R  
)_{\rm I}
\oplus
n_{\nu_{{\rm I},R}} {\bar{\nu}_{{\rm I},R}}
\sim 
\big((\overline{\bf 3},{\bf 1})_{2} \oplus ({\bf 1},{\bf 2})_{-3}  
\oplus
({\bf 3},{\bf 2})_{1} \oplus (\overline{\bf 3},{\bf 1})_{-4} \oplus ({\bf 1},{\bf 1})_{6} \big)_{\rm I}
\oplus n_{\nu_{{\rm I},R}} {({\bf 1},{\bf 1})_{0}}
\quad
\quad
\eea
for each family (family index ${\rm I},{\rm J}=1,2,3$; 
with ${\psi_L}_1$ for $u,d,e$ type,
${\psi_L}_2$ for $c,s,\mu$ type,
and 
${\psi_L}_3$ for $t,b,\tau$ type of quarks and leptons)
of $su(3) \times su(2) \times u(1)_{\tilde Y}$.
We use ${\rm I}=1,2,3$ for $n_{\nu_{e,R}}, n_{\nu_{\mu,R}}, n_{\nu_{\tau,R}} \in \{ 0, 1\}$
to label either the absence or presence of electron $e$, muon $\mu$, or tauon $\tau$ types of sterile neutrinos.
{Readers should keep in mind that all unitary internal symmetries that we will discuss below are part of the subgroup of the largest internal $\U(15 N_f)$ or $\U(16 N_f)$ acting on those Weyl fermions as unitary rotations.} 
The SM lagrangian consists of Yang-Mills terms (their gauge sector indices ${I}=1,2,3$ for $u(1),su(2),su(3)$), 
a possible theta term for $su(3)$, 
the Weyl fermions coupled to Yang-Mills gauge fields,
Yukawa-Higgs term, and electroweak Higgs kinetic-potential term:
\bea
 \label{eq:LSM}
\cL_{\rm SM}
= {\sum_{{ I}=1,2,3}}
 -\frac{1}{4} F_{{I},\mu\nu}^\ra F_{{I}}^{\ra \mu\nu}
 {-\, \frac{{\theta_3}}{{64} \pi^2} {g}_3^2 \epsilon^{\mu\nu \mu'\nu'} F_{3,\mu\nu}^\ra F_{3, \mu'\nu'}^\ra}
+ {\psi}^\dagger_L  (\ii \bar  \sigma^\mu {D}_{\mu, A} ) \psi_L
 -( {\psi}^\dagger_L \phi \psi_R +{\rm h.c.})
+ | {D}_{\mu, A}\phi |^2 
-{\rm U}(\phi)
.\quad
\eea
The $\cL_{\rm YH}={\psi}^\dagger_L \phi \psi_R+{\rm h.c.}$ is a shorthand of
$\cL_{\rm YH}^d
+\cL_{\rm YH}^u
+\cL_{\rm YH}^e=\lambda^{d}_{\rm IJ} {{q}^{\rm I \dagger}_L} \phi d_R^{\rm J} 
  +\lambda^{u}_{\rm IJ} \epsilon^{ab} {{q}^{\rm I \dagger}_{L a}}\phi_b^* u_R^{\rm J}
  +\lambda^{e}_{\rm IJ} {{l}^{\rm I \dagger}_L} \phi e_R^{\rm J}+{\rm h.c.}$
  with $a,b$ the $su(2)$ fundamental's index, and the h.c. as hermitian conjugate.
Diagonalization of Yukawa-Higgs term of quark sector implies {that} the $W^{\pm}$ boson induces a flavor-changing current mixing between different families,
thus we only have a $\U(1)_{\bf B}$ symmetry for all quarks (instead of {an} individual $\U(1)$ for each quark family), at least \emph{classically}.
The diagonalization of Yukawa-Higgs term of lepton sector without neutrino mass term
$\cL_{\rm YH}^{\nu}=\lambda^{\nu}_{\rm IJ} \epsilon^{ab} {{l}^{\rm I \dagger}_{L a}}\phi_b^* \nu_R^{\rm J}+{\rm h.c.}$
implies that $\cL_{\rm SM}$ has individual $\U(1)_{e}$, $\U(1)_{\mu}$, $\U(1)_{\tau}$ for each lepton family.
However, established experiments say each lepton $\U(1)$ is violated, 
only the total lepton number $\U(1)_{\bf L}$ should be considered, at least \emph{classically}.
Thus, we focus on $\U(1)_{\bf B}$ and $\U(1)_{\bf L}$
transformations, which send 
\bea
\U(1)_{\bf B}: ({\psi_L})_{\rm I}  &\mapsto&
((\e^{-\ii \frac{\alpha_{\bf B}}{3}} \mathbb{I}_3 \cdot \bar{d}_R) \oplus
 {l}_L  \oplus 
(\e^{\ii \frac{\alpha_{\bf B}}{3}} \mathbb{I}_6  \cdot q_L)  \oplus 
(\e^{-\ii \frac{\alpha_{\bf B}}{3}} \mathbb{I}_3  \cdot \bar{u}_R) \oplus   \bar{e}_R 
)_{\rm I}
\oplus
n_{\nu_{{\rm I},R}} {\bar{\nu}_{{\rm I},R}},\cr
\U(1)_{\bf L}: 
({\psi_L})_{\rm I}  &\mapsto&
(\bar{d}_R \oplus  (\e^{\ii {\alpha_{\bf L}}} \mathbb{I}_2 \cdot {l}_L)  \oplus q_L  \oplus \bar{u}_R \oplus   (\e^{-\ii {\alpha_{\bf L}}}\bar{e}_R)
)_{\rm I}
\oplus
 (\e^{-\ii {\alpha_{\bf L}}} n_{\nu_{{\rm I},R}} {\bar{\nu}_{{\rm I},R}}),
\eea 
with ${\alpha_{\bf B}} \in [0, 2 \pi \cdot 3)$ and ${\alpha_{\bf L}} \in [0, 2 \pi)$. 
The quark's $\U(1)_{q}$ is related to baryon's $\U(1)_{\bf B}$ via ${\alpha_{q}} ={\alpha_{\bf B}}/3 \in [0, 2 \pi)$.
Here $\mathbb{I}_{\rm N}$ means a rank-N diagonal identity matrix that can act on the N-multiplet.

It is well-known that the 
$\U(1)_{\bf B}$-$\U(1)_{\tilde Y}^2$, 
$\U(1)_{\bf L}$-$\U(1)_{\tilde Y}^2$ 
$\U(1)_{\bf B}$-$\SU(2)^2$ and 
$\U(1)_{\bf L}$-$\SU(2)^2$ have 
local anomalies captured by triangle Feynman diagrams, with their anomaly coefficients respectively:
$2 \cdot 1^2 -2^2 -(-4)^2=-18$, \;
$2 \cdot (-3)^2 -6^2=-18$,
1, and 1.
Upon dynamically gauging electroweak $su(2) \times u(1)_{\tilde Y}$,
the consequential ABJ anomaly implies that the classical continuous
$\U(1)_{\bf B} \times \U(1)_{\bf L}$ symmetry is broken \emph{quantum mechanically}.
Next we check whether any subgroup of the $\U(1)_{\bf B} \times \U(1)_{\bf L}$ still survives under dynamically gauged $G_{\SM_q}$.
The Fujikawa path integral method \cite{FujikawaSuzukiBook2004} shows that under $\U(1)_{\bf B}$ and $\U(1)_{\bf L}$ transformations
with corresponding currents $J^{}_{\bf B}$ and $J^{}_{\bf L}$,
the path integral $\bf Z$ changes to
\bea \label{eq:instanton-breaking}
\int [\cD \psi_L][\cD \psi_L^\dagger]
\e^{\ii \big( 
\int\dd^4x \big( \cL_{\rm SM}
+\alpha_{\bf B}(
\prt_{\mu} J^{\mu}_{\bf B}
)
+\alpha_{\bf L}(
\prt_{\mu} J^{\mu}_{\bf L}
)
\big)
{ {-18(\alpha_{\bf B}+\alpha_{\bf L}) N_f} n^{(1)} }
{- {(\alpha_{\bf B}+\alpha_{\bf L})N_f} n^{(2)}}
\big) 
}.
\eea
Here the instanton numbers
$n^{(1)} \equiv  \int\dd^4x \frac{{g}_1^2}{{32} \pi^2}  \epsilon^{\mu\nu \mu'\nu'} F_{1,\mu\nu} F_{1, \mu'\nu'}$
and 
$n^{(2)} \equiv  \int\dd^4x \frac{{g}_2^2}{{64} \pi^2}  \epsilon^{\mu\nu \mu'\nu'} F_{2,\mu\nu} F_{2, \mu'\nu'}$
are quantized in integer $\Z$ on spin manifolds.

\noindent
$\bullet$ $\U(1)_{{\bf B} - {\bf L}}$ symmetry: 
When $\alpha_{\bf B - L} \equiv \alpha_{\bf B} = -\alpha_{\bf L}$, 
its Ward identity (the derivative of the partition function ${\bf Z}$ with respect to $\alpha$ variation 
$\frac{\delta{\bf Z}}{\delta (\alpha_{})} \big\vert_{\alpha =0}=0$ vanishes)
says that
$\langle\prt_{\mu} (J^{\mu}_{\bf B}- J^{\mu}_{\bf L} ) \rangle=0$.
This shows $\U(1)_{{\bf B} - {\bf L}}$ is still a symmetry and is anomaly free under $G_{\SM_q}$.

\noindent
$\bullet$ $\Z_{2N_f, \mathbf{B} + \mathbf{L}}$ symmetry:
The continuous $\U(1)_{{\bf B}}$, $\U(1)_{{\bf L}}$ and $\U(1)_{{\bf B}+  {\bf L}}$ all are broken by ABJ anomaly.
But when $\alpha_{\bf B + L} \equiv \alpha_{\bf B} = \alpha_{\bf L}$, 
the quantization ${ {18(\alpha_{\bf B}+\alpha_{\bf L}) N_f} n^{(1)} }
={ {18 \alpha_{\bf B + L} (2N_f}) n^{(1)} } \in 2 \pi \Z$ under U(1) instanton
holds when $\alpha_{\bf B + L}  \in \frac{2 \pi}{36 N_f} \Z$;
the quantization ${ {(\alpha_{\bf B}+\alpha_{\bf L}) N_f} n^{(2)} }
= \alpha_{\bf B + L} (2N_f) \in 2 \pi \Z$ under SU(2) instanton
holds when $\alpha_{\bf B + L}  \in \frac{2 \pi}{2 N_f} \Z$.
Overall, this shows $\Z_{2N_f, \mathbf{B} + \mathbf{L}}$  is still a symmetry and is anomaly free under $G_{\SM_q}$.

\noindent
$\bullet$ $\U(1)_{{\bf B} - {\bf L}} \times_{\Z_2^{F}} \Z_{2N_f, \mathbf{B} + \mathbf{L}}$ symmetry:
Since $\U(1)_{{\bf B} - {\bf L}}$ and $\Z_{2N_f, \mathbf{B} + \mathbf{L}}$ share the fermion parity $\Z_2^{F}$ that acts on fermions $\psi \mapsto -\psi$,
the precise surviving subgroup (which is not broken under ABJ anomaly with dynamically gauged $G_{\SM_q}$)
is $\U(1)_{{\bf B} - {\bf L}} \times_{\Z_2^{F}} \Z_{2N_f, \mathbf{B} + \mathbf{L}}$.
Hereafter 
we use a standard notation that
${{G_1} \times_{{G_N}}  {{G}_2 }} \equiv ({\frac{{G_{1}} \times  {{G}_{2} } }{{G_N}}})$ as modding out their common normal subgroup $G_N$.
 
The above result is not affected by whether the $({\psi_L})_{\rm I}$ multiplet in \eq{eq:SMrep} includes 
the 16th Weyl fermion ${\bar{\nu}_{{\rm I},R}} =  {({\bf 1},{\bf 1})_{0}}$
for each of the $N_f$ families, because ${\bar{\nu}_{{\rm I},R}}$ is sterile to $G_{\SM_q}$ gauge forces.

\section{Proton Stability}

We have shown $\U(1)_{{\bf B} - {\bf L}}$ and $\Z_{2N_f, \mathbf{B} + \mathbf{L}}$ are good symmetries, 
anomaly-free with respect to 
dynamically gauged $G_{\SM_q}$ for the SM. 
Precisely they send the multiplet $({\psi_L})_{\rm I}$ to 
(here ${\alpha_{\bf B - L}} \in [0, 2 \pi \cdot 3)$ while $\alpha_{\bf B + L}  \in \frac{2 \pi}{2 N_f} \Z$ is discrete):  
\bea
\U(1)_{\bf B -L} &:&  
((\e^{-\ii \frac{\alpha_{\bf B - L}}{3}} \mathbb{I}_3 \cdot \bar{d}_R) \oplus
(\e^{-\ii {\alpha_{\bf B - L}}} \mathbb{I}_2 \cdot {l}_L) \oplus 
(\e^{\ii \frac{\alpha_{\bf B - L}}{3}} \mathbb{I}_6  \cdot q_L)  \oplus 
(\e^{-\ii \frac{\alpha_{\bf B - L}}{3}} \mathbb{I}_3  \cdot \bar{u}_R) \oplus
(\e^{\ii {\alpha_{\bf B - L}}}\bar{e}_R))_{\rm I}
\oplus
 (\e^{\ii {\alpha_{\bf B - L}}} n_{\nu_{{\rm I},R}} {\bar{\nu}_{{\rm I},R}})
.\cr
\hspace{-5mm}
\Z_{2N_f, \mathbf{B} + \mathbf{L}}&:& 
((\e^{-\ii \frac{\alpha_{\bf B + L}}{3}} \mathbb{I}_3 \cdot \bar{d}_R) \oplus
(\e^{\ii {\alpha_{\bf B + L}}} \mathbb{I}_2 \cdot {l}_L) \oplus 
(\e^{\ii \frac{\alpha_{\bf B + L}}{3}} \mathbb{I}_6  \cdot q_L)  \oplus 
(\e^{-\ii \frac{\alpha_{\bf B + L}}{3}} \mathbb{I}_3  \cdot \bar{u}_R) \oplus
(\e^{-\ii {\alpha_{\bf B + L}}}\bar{e}_R))_{\rm I}
\oplus
 (\e^{-\ii {\alpha_{\bf B + L}}} n_{\nu_{{\rm I},R}} {\bar{\nu}_{{\rm I},R}})
. \cr
&&
\eea
Let us check whether they are symmetries for the familiar GUTs 
and UU. 
\begin{enumerate}[leftmargin=-0mm]
\item $\U(1)_{{\bf B} - {\bf L}}$ vector or $\U(1)_{X}$ chiral like symmetry:
$\U(1)_{{\bf B} - {\bf L}}$ is a factor of the left-right (LR) $su(3) \times su(2)_{\rL} \times su(2)_{\rR} \times u(1)_{\frac{ \mathbf{B}-  \mathbf{L}}{2}}$
model \cite{SenjanovicMohapatra1975}.
$\U(1)_{{\bf B} - {\bf L}}$ is part of $\SU(4)$ subgroup of Pati-Salam (PS) $su(4) \times su(2)_{\rL}  \times su(2)_{\rR} $ \cite{Pati1974yyPatiSalamLeptonNumberastheFourthColor}.
Thus $\U(1)_{{\bf B} - {\bf L}}$ is not only an anomaly-free symmetry but also already dynamically gauged for LR and PS models.
\cblue{Similarly for the Trinification (Tri) \cite{1979AchimanStechTrinification, 1984deRujulaGeorgiGlashow, BabuHePakvasa1985giTrinification}
with the gauged Lie algebra $su(3) \times su(3)_{\rL} \times su(3)_{\rR}$,
the $\U(1)_{{\bf B} - {\bf L}}$ is contained in the $\SU(3)_{\rR}$; 
thus $\U(1)_{{\bf B} - {\bf L}}$ is automatically anomaly-free and dynamically gauged in the Trinification.}
$\U(1)_{{\bf B} - {\bf L}}$ is not quite correct for Georgi-Glashow $su(5)$ \cite{Georgi1974syUnityofAllElementaryParticleForces},
because $\U(1)_{{\bf B} - {\bf L}}$ does not act on the 
${\psi_L}$ reducibly as $\overline{\bf 5}$, ${\bf 10}$, ${\bf 1}$ multiplets of $su(5)$.
But the $\U(1)_X$ with $X \equiv 5({ \mathbf{B}-  \mathbf{L}})-\frac{2}{3} {\tilde Y}$ \cite{Wilczek1979hcZee, WilczekZeePLB1979} fits the role so 
that $(\bar{d}_R \oplus {l}_L ) \oplus (q_L  \oplus \bar{u}_R \oplus   \bar{e}_R) \oplus ( \bar{\nu}_R) =\overline{\bf 5}_{-3} \oplus {\bf 10}_{1} \oplus {\bf 1}_{5}$
in $su(5) \times u(1)_X$. 
Neither $\U(1)_{{\bf B} - {\bf L}}$ nor $\U(1)_X$ is a symmetry for the flipped $u(5)$ model \cite{Barr1982flippedSU5},
but the $\U(1)_{X_2}$
with $X_2 \equiv \frac{1}{5} X + \frac{4}{5} \tilde{Y} = ({ \mathbf{B}-  \mathbf{L}}) + \frac{2}{3} {\tilde Y}$
replaces the role so 
that $(\bar{u}_R \oplus {l}_L ) \oplus (q_L  \oplus \bar{d}_R \oplus   \bar{\nu}_R) \oplus ( \bar{e}_R) =\overline{\bf 5}_{-3} \oplus {\bf 10}_{1} \oplus {\bf 1}_{5}$
in $su(5) \times u(1)_{X_2}$. 
Neither $\U(1)_{{\bf B} - {\bf L}}$, $\U(1)_{X}$, nor $\U(1)_{X_2}$
is correct for the $so(10)$ GUT \cite{Fritzsch1974nnMinkowskiUnifiedInteractionsofLeptonsandHadrons}.
The only sensible U(1) factor allowed for the $so(10)$ GUT is an identical $\U(1)$ phase that acts on the ${\bf 16}$
that together with Spin(10) forms a $(\Spin(10) \times_{\Z_{4,X}} \U(1))$ internal symmetry. 
But this chiral $\U(1)$ has a
discrete ${\Z_{4,X}} ={\Z_{4,X_2}}$ subgroup 
(that replaces the role of discrete ${{\bf B} - {\bf L}}$) which is a good symmetry (indeed anomaly-free and gauged) for the $so(10)$ GUT.
Finally, UU \cite{JW2006.16996, JW2008.06499, JW2012.15860} 
requires a discrete ${{\bf B} - {\bf L}}$ or $X$ symmetry to enforce a $\Z_{16}$ class global anomaly constraint
that can be canceled by replacing the $\bar{\nu}_R$ with TQFT/CFT exotic phases.
See \eq{eq:B-L-symmetry} for a summary.

\item $\Z_{2N_f, \mathbf{B} + \mathbf{L}}$ vector symmetry:
$\Z_{2N_f, \mathbf{B} + \mathbf{L}}$ is a good global symmetry for LR model,
broken down from $\U(1)_{{\bf B} + {\bf L}}$ 
by the dynamical $\SU(2)_{\rL}$ and $\SU(2)_{\rR}$ instantons.
For other models such as PS, $su(5)$, flipped $u(5)$ and $so(10)$,
the only compatible subgroup of $\U(1)_{{\bf B} + {\bf L}}$ is the fermion parity $\Z_2^{F}$. 
For UU, we have choices to include the $15N_f$ Weyl fermion SM only but without GUT structure plus exotic TQFT/CFT sectors;
those UU models can obey the $\Z_{2N_f, \mathbf{B} + \mathbf{L}}$ symmetry.
Other UU which includes the $su(5)$ GUT allows only the $\Z_2^{F}$ subgroup of $\Z_{2N_f, \mathbf{B} + \mathbf{L}}$.
\cblue{For Trinification, we can even choose a generalized unbroken $\U(1)_{{\bf B} + {\bf L}}$ global symmetry, 
which is not only outside the Trinification gauge group 
but also anomaly-free under that gauge group with Lie algebra $su(3) \times su(3)_{\rL} \times su(3)_{\rR}$.}
See \eq{eq:B-L-symmetry} for a summary.
\end{enumerate}
\bea \label{eq:B-L-symmetry}
\begin{tabular}{c | c c c c c  c c c }
\hline
		 &  SM & LR \cite{SenjanovicMohapatra1975}\quad &  PS \cite{Pati1974yyPatiSalamLeptonNumberastheFourthColor} \quad & $su(5)$  \cite{Georgi1974syUnityofAllElementaryParticleForces} \quad
		  & flipped $u(5)$ \cite{Barr1982flippedSU5} \quad & $so(10)$ \cite{Fritzsch1974nnMinkowskiUnifiedInteractionsofLeptonsandHadrons} 
		  & \cblue{Tri} \cite{1979AchimanStechTrinification, 1984deRujulaGeorgiGlashow, BabuHePakvasa1985giTrinification}
		  & UU \cite{JW2012.15860}
		\\ 
		\hline
		 ${{\bf B} - {\bf L}}$ like & $\U(1)_{{\bf B} - {\bf L}}$ & $\U(1)_{{\bf B} - {\bf L}}$
		  &    $\U(1)_{{\bf B} - {\bf L}}$ &  $\U(1)_X$  & $\U(1)_{X_2}$ & ${\Z_{4,X}}$  &    \cblue{$\U(1)_{{\bf B} - {\bf L}}$}   & discrete ${{\bf B} - {\bf L}}$ or $X$\\
		  \hline
		 \cblue{${{\bf B} + {\bf L}}$ like} & $\Z_{2N_f, \mathbf{B} + \mathbf{L}}$  & $\Z_{2N_f, \mathbf{B} + \mathbf{L}}$
		  & $\Z_2^{F}$   &  $\Z_2^{F}$ & $\Z_2^{F}$ & $\Z_2^{F}$ & \cblue{$\U(1)_{{\bf B} + {\bf L}}$} & $\Z_{2N_f, \mathbf{B} + \mathbf{L}}$ or $\Z_2^{F}$ \\
		\hline
\end{tabular}.\quad\quad
\eea

We check whether these symmetries avoid the proton $p^+$ (or other nucleons like neutron $n$)
decay for some dominant channels.
These channels are constrained to have the lifetime $\tau_T$ lower bound around or more than $10^{33}$ years \cite{PDG2020, NathFileviez0601023, Senjanovic0912.5375}.
We list down the changes of baryon or lepton number ($\Delta {\bf B}$ or $\Delta {\bf L}$):  
\bea \label{eq:proton}
\begin{tabular}{c | c | c |  c | c  | c | c}
\hline
& $p^+ \to e^+ \pi^0$ & $p^+ \to \mu^+ \pi^0$ & $p^+ \to \mu^+ K^0$ & $p^+ \to e^+ K^0$ & $n \to e^+ K^-$ & $n \to e^- K^+$ \\
\hline
$(\Delta {\bf B}, \Delta {\bf L})$  & $(-1,-1)$  &  $(-1,-1)$   &  $(-1,-1)$  &  $(-1,-1)$  &  $(-1,-1)$  &  $(-1,+1)$ \\
\hline
\end{tabular}.
\eea
All these processes have $\Delta ({\bf B} - {\bf L}) =0$ and $\Delta ({\bf B} + {\bf L}) =-2$,
except the last one has $\Delta ({\bf B} - {\bf L}) =-2$ and $\Delta ({\bf B} + {\bf L}) =0$
If $\Z_{2N_f, \mathbf{B} + \mathbf{L}}$ is an exact symmetry of our vacuum, 
then $\Delta ({\bf B} + {\bf L}) = 0 \rmod 2N_f = 0 \rmod 6$ must hold. Many proton decay channels are thus forbidden.
The last channel $n \to e^- K^+$ is also unlikely because it violates 
even the discrete $\Z_{4, \mathbf{B} - \mathbf{L}}$ which demands that $\Delta ({\bf B} - {\bf L}) =0 \mod 4$.

Proton decay is forbidden as long as $\Z_{2N_f, \mathbf{B} + \mathbf{L}}$ is an exact symmetry and $N_f >1$.
\ccblue{(Although a single proton decay may be forbidden, 
the $\Z_{2N_f, \mathbf{B} + \mathbf{L}}$ symmetry still allows $N_f$ protons or baryons together to 
decay because their $\Delta ({\bf B} + {\bf L}) =-2 \times N_f = - 6 =0 \rmod 6$ is consistent with $\Z_{2N_f, \mathbf{B} + \mathbf{L}}$
here for $N_f=3$.)}
If all experiments support that \ccblue{a proton} indeed does not decay at all in our vacuum, then the SM and the UU (that contains only SM without GUT) 
are viable candidates deserving further studies, 
their generalizations with $\Z_{2N_f, \mathbf{B} + \mathbf{L}}$ symmetry is preferable for future model buildings.
\ccblue{For example, anomaly-free discrete gauge symmetries were recently discussed in \cite{Davighi2202.05275}
(see also references therein), which serve to exactly stabilize the
proton via the all-orders selection rule.}

%

\section{Cobordism 
and Higher Anomalies of the Standard Model with $\Z_{2N_f, \mathbf{B} + \mathbf{L}}$ symmetry
}

We have checked that $\U(1)_{{\bf B} - {\bf L}} \times_{\Z_2^{F}} \Z_{2N_f, \mathbf{B} + \mathbf{L}}$ 
is free from perturbative local anomaly under dynamical SM gauge group $G_{\SM_q} \equiv \frac{\SU(3) \times   \SU(2) \times \U(1)}{\Z_q}$.
Now we follow the procedure in \cite{WW2019fxh1910.14668, WangWanYou2112.14765}
to check whether the full spacetime-internal symmetry $G$ has any local and global anomalies, including all gauge or gravitational background fields.
The spacetime symmetry is Spin group (the double cover of the special orthogonal SO group), 
and the internal symmetry is $\U(1)_{\mathbf{B} - \mathbf{L}} \times_{\Z_2^F} \Z_{2N_f,  \mathbf{B} + \mathbf{L}} \times G_{\SM_q}$,
or $\Z_{4,X} \times_{\Z_2^F} \Z_{2N_f,  \mathbf{B} + \mathbf{L}} \times G_{\SM_q}$ if the discrete $X$ replacing the continuous ${\mathbf{B} - \mathbf{L}}$ symmetry.
We focus on $N_f=3$ thus $\Z_{6, \mathbf{B} + \mathbf{L}}= {\Z_2^F} \times \Z_{3, \mathbf{B} + \mathbf{L}}$.
Given the spacetime-internal $G$, we classify 
all possible 4d local ($\Z$ class) and global ($\mathbb{Z}_{\rm n}$ class) anomalies via computing the 
5th cobordism group $\TP_5(G)$ \cite{2016arXiv160406527F}:
\bea
&& \label{eq:TP5-Spin-SM-U1-Z3B+L}
\TP_5(\Spin \times_{\Z_2^F} \U(1)_{\mathbf{B} - \mathbf{L}} \times \Z_{3, {\mathbf{B} + \mathbf{L}}} \times G_{\SM_q})
=\big( \Z^{11} \big) \times \big(\Z_9 \times { \Z_3^7}\big).
 \\
 && \label{eq:TP5-Spin-SM-Z4-Z3B+L}
  \TP_5(\Spin \times_{\Z_2^F} \Z_{4,X} \times \Z_{3, {\mathbf{B} + \mathbf{L}}} \times G_{\SM_q})
=\left\{\begin{array}{ll}\big( \Z^5  \times\Z_2\times\Z_4^2 \times \Z_{16}\big) \times \big( \Z_9 \times  \Z_3^4 \big),&q=1,3.\\
\big(\Z^5  \times\Z_2^2\times\Z_4 \times\Z_{16} \big)\times \big(\Z_9 \times  \Z_3^4\big),&q=2,6.
\end{array}
\right.
\eea
The anomaly classifications on the right-hand side in the first big bracket $\big(\dots\big)$ are obtained in \cite{WW2019fxh1910.14668, WangWanYou2112.14765}, 
while those in the second big bracket $\big(\dots\big)$ {are} new to the literature.
The $\Z^{11}$ and $\Z^5$ classes of local anomalies are familiar to QFT textbook readers \cite{Weinberg1996Vol2}.
The $\Z_2^{...}\times\Z_4^{...} \times \Z_{16}$ classes of global anomalies were characterized in detail before \cite{WW2019fxh1910.14668}.
These anomalies all canceled, except only $\Z^{2}$ (in $\Z^{11}$) classes (involving the cubic pure gauge $\U(1)_{{ \mathbf{B}-  \mathbf{L}}}^3$ and 
mixed gauge-gravity $\U(1)_{{ \mathbf{B}-  \mathbf{L}}}$-(gravity)$^2$ anomalies) and the $\Z_{16}$ class 
(involving the mixed gauge-gravity global anomaly between the $\Z_{4,X}$ and the spacetime diffeomorphism), 
anomalies can have a nonzero coefficient
$-N_f+n_{\nu_{R}} \equiv 
-N_f+
\sum_{\rm I} 
 n_{\nu_{{\rm I},R}}$ if the number of family $N_f$ is distinct from the total number of right-handed neutrinos $n_{\nu_{R}}$ \cite{WangWanYou2112.14765}.
The nonzero $\Z^2$ and $\Z_{16}$ anomalies of ${\mathbf{B}-  \mathbf{L}}, X$ and gravity background fields
as 't Hooft anomalies
do not imply the sickness of SM  as long as ${\mathbf{B}-  \mathbf{L}}, X$ and gravity are \emph{non-dynamical}.
But when ${\mathbf{B}-  \mathbf{L}}, X$ and gravity are \emph{dynamical},
to cancel these anomalies \cite{McNamara2019rupVafa1909.10355}, 
we can either 
introduce enough $n_{\nu_{R}}$, or break ${\mathbf{B}-  \mathbf{L}}$, or introduce new exotic phases from UU \cite{JW2006.16996, JW2008.06499, JW2012.15860}. 

The $\Z_9 \times  {\Z_3^7}$ global anomalies in \eq{eq:TP5-Spin-SM-U1-Z3B+L}
(which exactly overlaps the $\Z_9 \times  \Z_3^4$ in \eq{eq:TP5-Spin-SM-Z4-Z3B+L} for the first five generators),
involving the discrete $\mathbf{B}-\mathbf{L}$ background field, are new to the literature: 
they are generated by cobordism invariants
{$\mathfrak{P}_3(B'_{\Z_{3,\mathbf{B} + \mathbf{L}}})$,
$A'_{\Z_{3,\mathbf{B} + \mathbf{L}}}B'_{\Z_{3,\mathbf{B} + \mathbf{L}}}c_1(\U(1)),A'_{\Z_{3,\mathbf{B} + \mathbf{L}}}c_1(\U(1))^2$,
$A'_{\Z_{3,\mathbf{B} + \mathbf{L}}}c_2(\SU(2))$,
$A'_{\Z_{3,\mathbf{B} + \mathbf{L}}}c_2(\SU(3))$,
$A'_{\Z_{3,\mathbf{B} + \mathbf{L}}}B'_{\Z_{3,\mathbf{B} + \mathbf{L}}}c_1(\U(1)_{\mathbf{B} - \mathbf{L}})$,
$A'_{\Z_{3,\mathbf{B} + \mathbf{L}}}c_1(\U(1)_{\mathbf{B} - \mathbf{L}})^2$, 
and {$A'_{\Z_{3,\mathbf{B} + \mathbf{L}}}c_1(\U(1)_{\mathbf{B} - \mathbf{L}})c_1(\U(1))$}.
}
Here the generators are written for the $q=1$ case.
Let us explain the notations.
Here all cohomology classes are pulled back to the manifold $M$ along the maps given in the definition of cobordism groups, e.g.
the $A'_{\Z_{3,\mathbf{B} + \mathbf{L}}} \in \H^1(\B{\Z_{3,\mathbf{B} + \mathbf{L}}}, \Z_3)$ is pulled back to $\H^1(M, \Z_3)$.
The $B'_{\Z_{3,\mathbf{B} + \mathbf{L}}} \equiv 
(\frac{1}{3} \delta A'_{\Z_{3,\mathbf{B} + \mathbf{L}}} \rmod 3)   
\in \H^2(\B{\Z_{3,\mathbf{B} + \mathbf{L}}}, \Z_3)$ is pulled back to $\H^2(M, \Z_3)$,
{defined} via the coboundary operator $\frac{1}{3} \delta$ for cochains (similar to the differential operator $\dd$ for differential forms) that maps $\H^1$ to $\H^2$.
The {$c_j({\rm G})$} is the $j$th Chern class of the associated vector bundle {of the principal G-bundle}.
The Postnikov square $\mathfrak{P}_3(B'_{\Z_{3,\mathbf{B} + \mathbf{L}}}) \equiv 
\beta_{(9,3)}(B'_{\Z_{3,\mathbf{B} + \mathbf{L}}} \cup B'_{\Z_{3,\mathbf{B} + \mathbf{L}}})$ maps $\H^2(...,\Z_{3})\to\H^5(...,\Z_{9})$
via the cup product $\cup$ for cochains (similar to the wedge product $\wedge$ for differential forms)
and the Bockstein homomorphism $\beta_{(9,3)}$ 
associated with the induced long exact sequence of cohomology with coefficients from the 
short exact sequence $0\to\Z_{9}\to\Z_{27}\to\Z_{3}\to 0$.
Cobordism invariants for other $q$ are similar \cite{WW2019fxh1910.14668}.  
For $q=2$, the $\SU(2)$ and $\U(1)$ labels change to $\U(2)$. 
For $q=3$, the $\SU(3)$ and $\U(1)$ labels change to $\U(3)$. 
For $q=6$, the $\SU(2)$ label changes to $\U(2)$, the $\SU(3)$ label changes to $\U(3)$, 
and the $\U(1)$ label changes to either $\U(2)$ or $\U(3)$ since $c_1(\U(2))=c_1(\U(3))$. 

Because all these $\Z_9 \times { \Z_3^7}$ global anomalies involve internal symmetries (thus gauge anomalies among $\mathbf{B} \pm \mathbf{L}$, $X$, and $G_{\SM_q}$)
without turning on the spacetime symmetry background (thus they are not mixed gauge-gravitational or gravitational anomalies), 
as long as these internal symmetries are dynamically gaugeable, we expect all these anomalies are canceled.

Moreover, once $G_{\SM_q}$ is dynamically gauged as in the SM vacuum,
we obtain extra 1-form electric and magnetic symmetries, $G_{[1]}^e$ and $G_{[1]}^m$ \emph{kinematically} 
as generalized global symmetries \cite{Gaiotto2014kfa1412.5148}
that act on charged objects (1d electric Wilson lines and 1d magnetic 't Hooft lines).
By gauging $G_{\SM_q}$, we obtain
$G_{[1]}^e \times G_{[1]}^m = \Z_{6/q,[1]}^e \times \U(1)_{[1]}^m$ 
\cite{Wan2019sooWWZHAHSII1912.13504, AnberPoppitz2110.02981, WangYou2111.10369GEQC}
in the dynamical SM gauge theory, for $q=1,2,3,6$ and we define $q' \equiv 6/q \equiv {2^{n_2} \cdot 3^{n_3}}=6,3,2,1$ as
$(n_2,n_3)=(1,1), (0,1), (1,0), (0,0)$ respectively.
We compute the potential 't Hooft anomaly classification of the gauged-$G_{\SM_q}$ SM 
via removing $G_{\SM_q}$ and including $G_{[1]}^e \times G_{[1]}^m$ into \eq{eq:TP5-Spin-SM-U1-Z3B+L} and 
\eq{eq:TP5-Spin-SM-Z4-Z3B+L}, 
\bea
&&  \label{eq:TP5-Spin-SM-U1-Z3B+L-higher}
\TP_5(\Spin \times_{\Z_2^F} \U(1)_{\mathbf{B} - \mathbf{L}} \times   \Z_{3, {\mathbf{B} + \mathbf{L}}}  \times \Z_{6/q,[1]}^e \times \U(1)_{[1]}^m)
=\big(\Z^2  \times \Z_{6/q}\big)  \times \big( \Z_9  \times  (\Z_3)^{2} \times  {(\Z_3)^{3 n_3} } \big).\\
 &&  \label{eq:TP5-Spin-SM-Z4-Z3B+L-higher}
 \TP_5(\Spin \times_{\Z_2^F} \Z_{4,X} \times   \Z_{3, {\mathbf{B} + \mathbf{L}}}  \times \Z_{6/q,[1]}^e \times \U(1)_{[1]}^m)
=
\big(\Z_{16}  \times (\Z_4)^{n_2} \times  \Z_{6/q}\big) \times \big( \Z_9 \times  (\Z_3)^{2 n_3} \big)
.
\eea
The anomaly classifications on the right-hand side in the first big bracket $\big(\dots\big)$ are obtained recently in \cite{WangWanYou2112.14765}, 
while those in the second big bracket $\big(\dots\big)$ {are} new to the literature.
The $\Z^2$ local and $\Z_{16}$ global anomalies are discussed earlier 
in \eq{eq:TP5-Spin-SM-U1-Z3B+L} and \eq{eq:TP5-Spin-SM-Z4-Z3B+L}, 
with its anomaly coefficient proportional to $(-N_f+n_{\nu_{R}})$ \cite{WangWanYou2112.14765}.
The $\Z_{6/q}$ global anomaly is a mixed anomaly between 1-form symmetries of $\Z_{6/q,[1]}^e$ and $\U(1)_{[1]}^m$, which is identified to be nonzero in \cite{WangWanYou2112.14765} --- which has dynamical constraints on the SM gauge theories. 
The $(\Z_4)^{n_2}$ global anomaly is canceled \cite{WangWanYou2112.14765}.
The $\Z_9$ global anomaly $\mathfrak{P}_3(B'_{\Z_{3,\mathbf{B} + \mathbf{L}}})$ and
the $(\Z_3)^{2}$ global anomalies $A'_{\Z_{3, {\mathbf{B} + \mathbf{L}}}} B'_{\Z_{3, {\mathbf{B} + \mathbf{L}}}}c_1(\U(1)_{\mathbf{B} - \mathbf{L}})$
and
$A'_{\Z_{3, {\mathbf{B} + \mathbf{L}}}}c_1(\U(1)_{\mathbf{B} - \mathbf{L}})^2$
are already canceled to zero in the SM earlier in \eq{eq:TP5-Spin-SM-U1-Z3B+L}.
The remaining {$(\Z_3)^{3 n_3}$ in \eq{eq:TP5-Spin-SM-U1-Z3B+L-higher}
(which exactly overlaps the $(\Z_3)^{2n_3}$ in \eq{eq:TP5-Spin-SM-Z4-Z3B+L-higher} for the first two generators)}
contains {three} more extra generators (for $n_3=1$):
$A'_{\Z_{3, {\mathbf{B} + \mathbf{L}}}} (B^e_{\Z_{3,[1]}})^2 $
,
$A'_{\Z_{3, {\mathbf{B} + \mathbf{L}}}} B'_{\Z_{3, {\mathbf{B} + \mathbf{L}}}} B^e_{\Z_{3,[1]}} $
and {$A'_{\Z_{3, {\mathbf{B} + \mathbf{L}}}}c_1(\U(1)_{\mathbf{B} - \mathbf{L}}) B^e_{\Z_{3,[1]}}$}.
{The first two of these three terms} are mixed anomalies 
between the 0-symmetry ${\mathbf{B} + \mathbf{L}}$ (coupled to 1-form background field $A'_{\Z_{3, {\mathbf{B} + \mathbf{L}}}} $)
and the electric 1-symmetry ${\Z_{3,[1]}}$ (coupled to 2-form background field $B^e_{\Z_{3,[1]}}$).
It is interesting to check whether these two mixed $({\mathbf{B} + \mathbf{L}})$-${\Z_{3,[1]}}$ anomalies 
occur for SM with $q=1,2$ (thus $n_3=1$). 
We can check this mixed anomaly between 0-symmetry and 1-symmetry 
via the techniques of the analogous anomaly studied previously in 
\cite{Gaiotto2017yupZoharTTT1703.00501,
Shimizu1706.06104, Tanizaki1711.10487, Cordova2018acb1806.09592DumitrescuClay, Wan2019oyr1904.00994, Wan2019oaxWWHAHSIII1912.13514}.
If the $\Z_3$ class of $A'_{\Z_{3, {\mathbf{B} + \mathbf{L}}}} (B^e_{\Z_{3,[1]}})^2$ anomaly is present, 
then this means that the $\Z_{2N_f, \mathbf{B} + \mathbf{L}} = \Z_{6, \mathbf{B} + \mathbf{L}}$ symmetry 
should be broken down to a $\Z_{2}^{F}$ subgroup
by the $\PSU(3)=\SU(3)/\Z_3$ fractional instanton.
Note that the instanton number
$n^{(N)} \equiv  \int\dd^4x \frac{{g}_2^2}{{64} \pi^2}  \epsilon^{\mu\nu \mu'\nu'} F_{\mu\nu} F_{ \mu'\nu'}$
for $\SU(N)$ gauge theory has $n^{(N)} \in \Z$ on spin manifolds,
while turning on 1-form symmetry ${\Z_{N,[1]}}$ background field 
can produce $\PSU(N)$ fractional instantons
such that $n^{(N)} \in \Z/N$ on spin manifolds, and $n^{(N)} \in \Z/N$ (for odd $N$) or $n^{(N)} \in \Z/2N$ (for even $N$) on non-spin manifolds.
So if $\SU(3)$ instanton 
was a source of $\U(1)_{\mathbf{B} + \mathbf{L}} \to \Z_{2N_f, \mathbf{B} + \mathbf{L}}$ symmetry breaking in \eq{eq:instanton-breaking},
the $\PSU(3)$ fractional instanton can further trigger the breaking 
$\Z_{2N_f, \mathbf{B} + \mathbf{L}} \to \Z_{2N_f/N_c, \mathbf{B} + \mathbf{L}} = \Z_{2}^{F}$ symmetry, here the family and color numbers {match} $N_f = N_c =3$.
But it is the $\SU(2)$ instanton 
(not the $\SU(3)$ instanton) causing the $\U(1)_{\mathbf{B} + \mathbf{L}} \to \Z_{2N_f, \mathbf{B} + \mathbf{L}}$ symmetry breaking.
So we conclude that the $\PSU(3)$ fractional instanton \emph{cannot} trigger the further breaking 
even in the presence of $\Z_{6/q,[1]}^e$-background field (for $q=1,2$).
Namely, this suggests no
$A'_{\Z_{3, {\mathbf{B} + \mathbf{L}}}} (B^e_{\Z_{3,[1]}})^2 $ anomaly in the gauged SM.

In comparison, \Refe{AnberPoppitz2110.02981} checked various fractional instanton contributions also showing a negative result on the 
mixed $({\mathbf{B} + \mathbf{L}})$-${\Z_{6/q,[1]}}$ anomalies. \Refe{AnberPoppitz2110.02981} found that 
in the presence of various fractional instantons constructed on a torus with 't Hooft twisted boundary condition \cite{tHooft1981nnxCMPHypertorus},
the continuous $\U(1)_{\mathbf{B} + \mathbf{L}}$ breaks down to 
$\Z_{2N_f, \mathbf{B} + \mathbf{L}}$, 
$\Z_{8N_f, \mathbf{B} + \mathbf{L}}$, 
$\Z_{4N_f, \mathbf{B} + \mathbf{L}}$, 
$\Z_{6N_f, \mathbf{B} + \mathbf{L}}$ for $q=1,2,3,6$.
But the familiar BPST SU(2) instanton \cite{BelavinBPST1975} already breaks $\U(1)_{\mathbf{B} + \mathbf{L}}$ 
to $\Z_{2N_f, \mathbf{B} + \mathbf{L}}$ shown in \eq{eq:instanton-breaking}.
Since $\Z_{2N_f, \mathbf{B} + \mathbf{L}}$ is not further broken down to its subgroup by fractional instantons,
\Refe{AnberPoppitz2110.02981} also found no mixed 't Hooft anomaly between $\Z_{2N_f, \mathbf{B} + \mathbf{L}}$ and ${\Z_{6/q,[1]}}$ symmetry.

{N\"aively, 
all these so-called 3-torsion $\big( \Z_9  \times  (\Z_3)^{2} \times  {(\Z_3)^{3 n_3} } \big)$ and $\big( \Z_9 \times  (\Z_3)^{2 n_3} \big)$
classes correspond to potential global anomalies in the SM 
between vector symmetries 
(the vector $\mathbf{B} \pm \mathbf{L}$ and the ${\Z^e_{3,[1]}}$ from gauging the vector $su(3)$ color), so there should be no anomalies among vector symmetries.
But the ${\Z^e_{3,[1]}}$ also partly descends from gauging the chiral $u(1)_{\tilde Y}$. 
In particular, the last two terms in the $(\Z_3)^{3 n_3}$, namely
$A'_{\Z_{3, {\mathbf{B} + \mathbf{L}}}} B'_{\Z_{3, {\mathbf{B} + \mathbf{L}}}} B^e_{\Z_{3,[1]}} \equiv 
A'_{\Z_{3, {\mathbf{B} + \mathbf{L}}}} (\frac{1}{3} \delta A'_{\Z_{3,\mathbf{B} + \mathbf{L}}} \rmod 3) B^e_{\Z_{3,[1]}}$ 
and $A'_{\Z_{3, {\mathbf{B} + \mathbf{L}}}}c_1(\U(1)_{\mathbf{B} - \mathbf{L}}) B^e_{\Z_{3,[1]}}$ anomalies, 
are subtle anomalies that cannot be checked by the fractional instanton argument above, thus they deserve further future investigations.}

\emph{Conclusion}
--- We had identified a continuous ${\mathbf{B} - \mathbf{L}}$ or discrete $X$, and discrete ${\mathbf{B} + \mathbf{L}}$ symmetry in the SM.
For $N_f=3$,
we had shown $\Z_{2N_f, \mathbf{B} + \mathbf{L}}$ is free from anomaly with SM gauged Lie group $G_{\SM_q}$ for all four versions of $q=1,2,3,6$.
We also classify all potential mixed anomalies between  $\Z_{2N_f, \mathbf{B} + \mathbf{L}}$ and gravitational background field (from Spin diffeomorphism)
or higher symmetries $\Z_{6/q,[1]}^e \times \U(1)_{[1]}^m$ of SM in \eq{eq:TP5-Spin-SM-U1-Z3B+L-higher} and 
\eq{eq:TP5-Spin-SM-Z4-Z3B+L-higher}. 
We show that $\Z_{2N_f, \mathbf{B} + \mathbf{L}}$ is free from all anomalies in \eq{eq:TP5-Spin-SM-U1-Z3B+L} and \eq{eq:TP5-Spin-SM-Z4-Z3B+L},
and it is free from many potential anomalies in \eq{eq:TP5-Spin-SM-U1-Z3B+L-higher} and 
\eq{eq:TP5-Spin-SM-Z4-Z3B+L-higher} involving higher symmetries as well. 
Although $\Z_{2N_f, \mathbf{B} + \mathbf{L}}$ is illegal in many GUTs in \eq{eq:B-L-symmetry}, 
$\Z_{2N_f, \mathbf{B} + \mathbf{L}}$ is a perfectly legal symmetry in the SM and some versions of UU to protect the proton stability.
Are there \emph{other ways} to protect the proton stability in UU? Notice that UU allows a 4d TQFT with discrete gauge forces (e.g., two layers of $[\Z_2]$ gauge theories) 
constructed out of the symmetry extension \cite{Wang2017locWWW1705.06728} (e.g., the first layer
$ 1 \to [\Z_2] \to \Spin \times {\Z_{4}} \to \Spin \times_{\Z_2^{F}} {\Z_{4,X}} \to 1$ and the second layer
$ 1 \to [\Z_2] \to \Spin \times {\Z_{8}} \to  \Spin \times {\Z_{4}}  \to 1$)
such that the even class of $\Z_{16}$ global anomaly in the original $\Spin \times_{\Z_2^{F}} {\Z_{4,X}}$ symmetry
is trivialized by pulled back to the extended $\Spin \times {\Z_{8}}$ symmetry \cite{Hsieh2018ifc1808.02881, JW2006.16996, JW2012.15860}.
Since the quark and baryon all carry odd ${\Z_{4,X}}$ charges,
it will be interesting to study the impact of 
the \emph{projective symmetry fractionalization} of $\Spin \times_{\Z_2^F} {\Z_{4,X}}$ on the line or surface operators 
of the 4d discrete gauge TQFT on the proton stability.

In the language of \emph{deformation class} of quantum field theories \cite{NSeiberg-Strings-2019-talk},
our present work together with \Refe{WangWanYou2112.14765} suggests that 
the 4d SM and GUTs can still be in the larger deformation class of QFT by including the $\Z_{2N_f, \mathbf{B} + \mathbf{L}}$ symmetry,
except that SM and UU can preserve $\Z_{2N_f, \mathbf{B} + \mathbf{L}}$, but many GUTs break $\Z_{2N_f, \mathbf{B} + \mathbf{L}}$ explicitly.
It will be important to understand deeper the implications of the subtle anomaly cancellation data presented in
\eq{eq:TP5-Spin-SM-U1-Z3B+L}, 
\eq{eq:TP5-Spin-SM-Z4-Z3B+L},
\eq{eq:TP5-Spin-SM-U1-Z3B+L-higher}, and 
\eq{eq:TP5-Spin-SM-Z4-Z3B+L-higher}, so we may use these results to investigate the SM dynamics better.

\emph{Acknowledgment}
---
JW thanks Yuta Hamada, Miguel Montero, and Cumrun Vafa for encouraging conversations on related topics.
During the preparation of this manuscript, we become aware that \Refe{KorenProtonStability2204.01741} by Seth Koren
also explores the relation between a discrete ${\mathbf{B} + \mathbf{L}}$ symmetry and the proton stability.
JW is supported by Center for Mathematical Sciences and Applications at Harvard University.
ZW is supported by the Shuimu Tsinghua Scholar Program. 
YZY is supported by a startup fund at UCSD.

\bibliography{BSM-Proton.bib}

\end{document}